\documentclass[12pt,a4paper]{article}
\usepackage{amsmath}
\usepackage{amssymb}
\usepackage{cite}

\renewcommand{\thesection}
  {\arabic{section}.\hspace{-.5em}}

\makeatletter
\renewcommand\section{
  \@startsection{section}{3}{\z@}%
  {-3.25ex\@plus -1ex \@minus -.2ex}%
  {1.5ex \@plus .2ex}%
  {\normalfont\normalsize\bfseries\mathversion{bold}}}
\renewcommand\subsection{
  \@startsection{subsection}{3}{\z@}%
  {-3.25ex\@plus -1ex \@minus -.2ex}%
  {1.5ex \@plus .2ex}%
  {\normalfont\normalsize\bfseries\mathversion{bold}}}
\makeatother

\makeatletter \@addtoreset{equation}{section} \makeatother
\renewcommand{\theequation}{\arabic{section}.\arabic{equation}}

\makeatletter
\renewcommand{\appendix}{
\renewcommand{\thesection}{\Alph{section}.\hspace{-.5em}}
\@addtoreset{equation}{section}
\renewcommand{\theequation}{\Alph{section}.\arabic{equation}}
\setcounter{section}{0}}
\makeatother

\newcommand{\Eqn}[1]{&\hspace{-0.5em}#1\hspace{-0.5em}&}
\newcommand{\nn}{\nonumber}
\renewcommand{\[}{\begin{equation}}
\renewcommand{\]}{\end{equation}}
\newcommand{\eqb}{\begin{eqnarray}}
\newcommand{\eqe}{\end{eqnarray}}

\newcommand{\grp}[1]{\mathrm{#1}}
\newcommand{\varth}{\vartheta}

\newcommand{\bbC}{{\mathbb C}}
\newcommand{\bbR}{{\mathbb R}}
\newcommand{\bbZ}{{\mathbb Z}}

\newcommand{\talpha}{{\tilde\alpha}}
\newcommand{\tbeta}{{\tilde\beta}}

\newcommand{\tvarphi}{{\tilde\varphi}}

\newcommand{\Rt}{R^\vee}
\newcommand{\bfR}{\boldsymbol{R}}

\newcommand{\pint}{\makebox[0pt][l]{\hspace{3.4pt}$-$}\int}

\begin{document}


\def\papertitlepage{\baselineskip 3.5ex \thispagestyle{empty}}
\def\preprinumber#1#2{\hfill
\begin{minipage}{1.2in}
#1 \par\noindent #2
\end{minipage}}

%
\papertitlepage
\setcounter{page}{0}
\preprinumber{arXiv:1312.1050}{}
\vskip 2ex
\vfill
\begin{center}
{\large\bf\mathversion{bold}
Thermodynamic limit of the Nekrasov-type formula\\[.5ex]
for E-string theory
}
\end{center}
\vfill
\baselineskip=3.5ex
\begin{center}
Takenori Ishii and Kazuhiro Sakai\\

{\small
\vskip 6ex
{\it Department of Physical Sciences, Ritsumeikan University}\\[1ex]
{\it Shiga 525-8577, Japan}\\
\vskip 1ex
{\tt gr0054vr@ed.ritsumei.ac.jp, ksakai@fc.ritsumei.ac.jp}

}
\end{center}
\vfill
\baselineskip=3.5ex
\begin{center} {\bf Abstract} \end{center}

We give a proof of the Nekrasov-type formula
proposed by one of the authors
for the Seiberg--Witten prepotential for
the E-string theory on $\bbR^4\times T^2$.
We take the thermodynamic limit of the Nekrasov-type formula
following the example of Nekrasov--Okounkov
and reproduce the Seiberg--Witten description of the prepotential.
The Seiberg--Witten curve obtained directly from the Nekrasov-type
formula is of genus greater than one. We find that this curve
is transformed into the known elliptic curve by a simple map.
We consider the cases in which the low energy theory has
$E_8,\,E_7\oplus A_1$ or $E_6\oplus A_2$ as a global symmetry.

\vfill
\noindent
December 2013


\setcounter{page}{0}
\newpage
\renewcommand{\thefootnote}{\arabic{footnote}}
\setcounter{footnote}{0}
\setcounter{section}{0}
\baselineskip = 3.5ex
\pagestyle{plain}
%

\section{Introduction}

The E-string theory is probably the simplest interacting
non-gravitational theory with (1,0) supersymmetry
in six dimensions \cite{Ganor:1996mu,Seiberg:1996vs}.
The theory is obtained as the low energy
theory of an M5-brane near one of the two fixed 9-planes
in the heterotic M-theory.
In the Coulomb branch it is a theory of
just one tensor multiplet.
Toroidal compactification of the theory exhibits rich structures
\cite{Klemm:1996hh,Minahan:1998vr,Ganor:1996pc}.
When compactified down to four dimensions,
the low energy effective theory is given by an ${\cal N}=2\ \grp{U}(1)$
gauge theory which is fully characterized by the Seiberg--Witten
solution \cite{Ganor:1996pc,Eguchi:2002fc,Eguchi:2002nx}.
See
\cite{Sakai:2011xg,Alim:2012ss,Klemm:2012sx,Haghighat:2013gba,
      Huang:2013yta}
for recent developments.

In \cite{Sakai:2012zq,Sakai:2012ik}
a Nekrasov-type formula for
the Seiberg--Witten prepotential for the E-string theory
was proposed.
It was also pointed out that 
the formula can be regarded as a special case of
the elliptic generalization of
a certain Nekrasov partition function
\cite{Nekrasov:2002qd,Hollowood:2003cv}.
The Nekrasov formula for ordinary gauge theories
has been proved \cite{Nekrasov:2003rj,Nakajima:2003pg}.
More specifically, it has been shown that the prepotential
obtained from the Nekrasov partition function is identical
to that prescribed in terms of the Seiberg--Witten curve.
It is natural to expect that the Nekrasov-type formula
for the E-string theory can be proved in a similar manner.
In this paper, we follow the example of
Nekrasov and Okounkov \cite{Nekrasov:2003rj} and give a proof
of the Nekrasov-type formula for the E-string theory.

It is important to note that 
interpretation of parameters in the Nekrasov-type formula
for the E-string theory is quite different from
the conventional one for ordinary gauge theories. For instance,
the parameter which represents the IR gauge coupling
in the case of conformal gauge theories is identified with
the expectation value of the Higgs field in the low energy theory of
the E-string theory. Because of this difference,
a straightforward generalization of the proof by
Nekrasov and Okounkov does not work for the E-string theory.
We overcome this problem by introducing
the antiderivative of the resolvent and deriving
Higgs expectation values from it.

Another nontrivial point is that
the Seiberg--Witten curve obtained directly from the Nekrasov-type
formula is of genus greater than one.
This is not a desired result because the known Seiberg--Witten
curve for the E-string theory is of genus one.
We resolve this mismatch by finding
a simple map which transforms the former higher genus curve
into the latter elliptic curve.

We mainly consider the simplest case, namely
the case where the E-string theory is compactified
on $T^2$ without Wilson line parameters.
In this case the low energy theory preserves
the original $E_8$ global symmetry.
The proof can be generalized to the cases with
nontrivial Wilson line parameters. As an illustration,
we discuss two examples which have
$E_7\oplus A_1$ or $E_6\oplus A_2$ as a global symmetry.

The organization of this paper is as follows.
In section~2, we review
the Seiberg--Witten description and the Nekrasov-type formula
for the E-string theory on $\bbR^4\times T^2$.
In section~3, we present our proof.
In section~4, we consider the cases where
$E_7\oplus A_1$ or $E_6\oplus A_2$
is a global symmetry.
Section~5 is devoted to the discussion.
In Appendix~A, we present the definitions of special functions
and some useful identities.

\section{Seiberg--Witten prepotential and Nekrasov-type formula}

In this section we review the Seiberg--Witten description of
the prepotential for the E-string theory on $\bbR^4\times T^2$
and its Nekrasov-type expression which we will prove
in the next section.

The low-energy theory of the E-string theory on $\bbR^4\times T^2$
is an ${\cal N}=2\ \grp{U}(1)$ gauge theory in four dimensions
and admits a Seiberg--Witten description
\cite{Seiberg:1994rs,Seiberg:1994aj}.
The low energy
effective action is fully characterized by a holomorphic function
called the prepotential. The prepotential can be expressed in terms of
the Seiberg--Witten curve. For the sake of simplicity here we
restrict ourselves to the simplest case
with trivial Wilson line parameters.
This is the case where the $E_8$ global symmetry is kept intact
under the torus compactification.
We will simply call it the $E_8$ theory hereafter.
Seiberg--Witten prepotential for this $E_8$ theory
was investigated in detail in \cite{Minahan:1997ct}.
See \cite{Sakai:2011xg} for detailed characterization
of the prepotential
including the cases with general Wilson line parameters.

The Seiberg--Witten curve for the $E_8$ theory is given by
\[\label{E8curve}
y^2=4x^3-\frac{1}{12}E_4(\tau)u^4x-\frac{1}{216}E_6(\tau)u^6+4u^5.
\]
Here $\tau$ is the complex modulus of the $T^2$ on which the E-string
theory is compactified and $u$ is the coordinate of
the Coulomb branch moduli space.
$E_{2n}(\tau)$ are the Eisenstein series (see Appendix A).
In the Seiberg--Witten description,
the expectation values of the Higgs fields in the vector multiplet
and the dual vector multiplet are respectively given by
\[\label{vevs}
\varphi(u,\tau)
=\frac{i}{4\pi^2}\int du \oint_\talpha\frac{dx}{y},\qquad
\varphi_{\rm D}(u,\tau)
=\frac{i}{4\pi^2}\int du \oint_\tbeta\frac{dx}{y}.
\]
Here one-cycles $\talpha,\tbeta$ of the elliptic curve (\ref{E8curve})
are chosen so that
\[\label{talphatbeta}
\oint_\talpha\frac{dx}{y}=\frac{2\pi}{u}
  +{\cal O}\left(\frac{1}{u^2}\right),\qquad
\oint_\tbeta\frac{dx}{y}=\frac{2\pi\tau}{u}
  +{\cal O}\left(\frac{1}{u^2}\right)
\]
for large $u$.
The integration constants of the integrals in $u$
are fixed in the way described below.
Inverting the function $\varphi(u,\tau)$ one can
express $\varphi_{\rm D}$
as a function in $\varphi$ and $\tau$.
The prepotential for the E-string theory is then prescribed as
\[\label{dF0inphi}
\frac{\partial F_0}{\partial\varphi}
  =8\pi^3i\left(\varphi_{\rm D}-\tau\varphi\right)+\mbox{const.}
\]
Here const.~could be a function in $\tau$,
but is a constant with respect to $\varphi$.
Throughout this paper we regard $\tau$ as a fixed parameter
rather than a variable.
Integrating the above expression
in $\varphi$, one obtains the prepotential.
The normalizations and the integration `constants'
of $\varphi,\varphi_{\rm D}$ and $F_0$ are fixed
so that the prepotential admits the following expansion
\[\label{F0series}
F_0(\varphi,\tau)=
\sum_{n=1}^\infty\sum_{k=0}^\infty N_{n,k}
\sum_{m=1}^\infty\frac{1}{m^3}e^{2\pi im(n\varphi+k\tau)}.
\]
Here $N_{n,k}$ are integers. 
The first few of them are \cite{Klemm:1996hh}
\begin{align}
\qquad\qquad&&N_{1,0}&=    1,
       &\quad N_{1,1}&=  252,
       &\quad N_{1,2}&= 5130,
       &\quad\cdots&\qquad\nn\\
\qquad\qquad&&N_{2,0}&=    0,
       &\quad N_{2,1}&=    0,
       &\quad N_{2,2}&=-9252,
       &\quad\cdots&\ .\qquad
\end{align}
This expression reflects the fact that the prepotential
can also be viewed as the BPS partition function of
the E-string theory on $\bbR^5\times S^1$
as well as the genus zero topological string amplitude
for the local $\tfrac{1}{2}$K3.
$N_{n,k}$ represent multiplicities of
BPS states in the effective theory in five dimensions
as well as those of rational curves in the local $\tfrac{1}{2}$K3.

Next, we recall the Nekrasov-type formula proposed in
\cite{Sakai:2012zq,Sakai:2012ik}.
Let us start with introducing some notations.
Let $E$ denote a two-dimensional torus $\bbC/(2\pi\bbZ+2\pi\tau\bbZ)$
and $\omega_k\ (k=0,1,2,3)$ be half periods of the torus:
\[
\omega_0=0,\qquad
\omega_1=\pi,\qquad \omega_2=-\pi-\pi\tau,\qquad \omega_3=\pi\tau.
\]
Throughout this paper the Weierstrass elliptic function $\wp(z)$
is defined over the torus $E$, i.e.~$\wp(z)=\wp(z;2\pi,2\pi\tau)$.
We let $\alpha, \beta$ denote two fundamental
one-cycles of the torus $E$.
They are chosen in such a way that
\[\label{alphabeta}
\oint_\alpha dz = 2\omega_1 = 2\pi,\qquad
\oint_\beta dz = 2\omega_3 = 2\pi\tau.
\]
Physically, $E$ may be understood as the dual torus
of the $T^2$ on which the E-string theory
is compactified. Here `dual' means that
Wilson line parameters with respect to
the directions in the $T^2$ take values on $E$.

Let $\bfR=(R_1,\ldots,R_N)$ denote an $N$-tuple
of partitions. Each partition $R_k$ is a nonincreasing sequence of 
nonnegative integers
\[
R_k = \{
  \mu_{k,1}\ge\mu_{k,2}\ge\cdots\ge\mu_{k,\ell(R_k)}>
  \mu_{k,\ell(R_k)+1}=\mu_{k,\ell(R_k)+2}=\cdots=0\}.
\]
Here the number of nonzero $\mu_{k,i}$ is denoted by $\ell(R_k)$. $R_k$
is represented by a Young diagram. We let $|R_k|$ denote the size of
$R_k$, i.e.~the number of boxes in the Young diagram of $R_k$:
\[
|R_k| := \sum_{i=1}^\infty\mu_{k,i} = \sum_{i=1}^{\ell(R_k)}\mu_{k,i}.
\]
Similarly, the size of $\bfR$
is denoted by
\[
|\bfR| := \sum_{k=1}^N |R_k|.
\]
We let $\Rt_k=\{\mu_{k,1}^\vee\ge\mu_{k,2}^\vee\ge\cdots\}$ denote the
conjugate partition of $R_k$. We also introduce the notation
\[
h_{k,l}(i,j):=\mu_{k,i}+\mu_{l,j}^\vee-i-j+1,
\]
which represents the relative hook-length of a box at $(i,j)$ between
the Young diagrams of $R_k$ and $R_l$.

We are now able to write down
the Nekrasov-type formula.
As discussed in \cite{Sakai:2012zq,Sakai:2012ik},
the formula can be expressed in several different ways.
For our present purposes it is convenient to express
the formula as a special case of
the elliptic generalization of the Nekrasov partition function
for the $\grp{U}(N)$ gauge theory
with $2N$ fundamental matters
\cite{Nekrasov:2002qd,Hollowood:2003cv}
\[\label{Zconformal}
Z:=\sum_{\bfR}
\left(-e^{2\pi i\varphi}\right)^{|\bfR|}
\prod_{k=1}^N
\prod_{(i,j)\in R_k}
\frac
{\prod_{n=1}^{2N}
 \varth_1\left(\tfrac{1}{2\pi}(a_k-m_n+(j-i)\hbar),\tau\right)}
{\prod_{l=1}^N
\varth_1\left(\tfrac{1}{2\pi}(a_k-a_l+h_{k,l}(i,j)\hbar),\tau\right)^2}.
\]
Here the sum is taken over all possible partitions $\bfR$
(including the empty partition).
A set of indices $(i,j)$ run over the coordinates of
all boxes in the Young diagram of $R_k$.
$\varth_1(z,\tau)$ is
the Jacobi theta function (see Appendix A).
For consistency we require
\[\label{amrel}
2\sum_{k=1}^N a_k - \sum_{n=1}^{2N} m_n = 0,
\]
where the equality should be regarded modulo periods of the torus $E$.
To obtain the prepotential for the E-string theory with four general
Wilson line parameters, we set
\[\label{fourWsetup}
N=4,\qquad
a_k=\omega_{k-1}\quad(k=1,2,3,4),\qquad
m_n=-m_{n+4}\quad(n=1,2,3,4).
\]
The Seiberg--Witten prepotential for the E-string theory
is then given by
\[\label{F0inZ}
F_0 = \left(2\hbar^2\ln Z\right)\big|_{\hbar=0}\,.
\]
The case of the $E_8$ theory is realized
by simply setting all the Wilson line
parameters $m_n$ to be zero. 
Actually, in this case one can simplify the expression
and express $Z$ as a sum over three partitions
\cite{Sakai:2012ik}.
More specifically, $Z$ for the $E_8$ theory
is given by (\ref{Zconformal}) with
\[\label{E8setup}
N=3,\qquad
a_k=\omega_k\quad(k=1,2,3),\qquad
m_n=0\quad(n=1,\ldots,6).
\]
We will use this simplified form in the proof below.

An important remark is that identification of parameters for
the E-string theory is quite different from what is known
for ordinary gauge theories. In the case of ordinary gauge
theories, $a_k$ represent diagonal elements of
the expectation value of the Higgs field
and $\varphi$ represents the IR gauge coupling. On the other hand, 
for the E-string theory $a_k$ are set to fixed values as above
and $\varphi$ represents the Higgs expectation value.
$\tau$ is the complex modulus of the $T^2$
and it plays the role of
the IR gauge coupling in the low energy theory in four dimensions.
Because of this difference,
a straightforward generalization of the proof by
Nekrasov--Okounkov \cite{Nekrasov:2003rj}
does not work in the case of E-string theory.
We will present a resolution to this problem in the next section.

\section{Proof}

In this section we prove that the prepotential given by
the Nekrasov-type formula in the last section is equivalent to
that expressed in terms of the Seiberg--Witten curve.
Our proof consists of three parts.
In subsection 3.1, we first take
the thermodynamic limit of the sum over partitions
(\ref{Zconformal}) and express the prepotential
as the solution of an extremum problem.
In subsection 3.2, we derive
the expression of the Higgs expectation value
in terms of the Seiberg--Witten curve.
In subsection 3.3,
we show that the prepotential obtained in the thermodynamic limit
is indeed equivalent to that expressed
in the Seiberg--Witten description.

In the proof below
we restrict ourselves to the case of the $E_8$ theory
and eventually set parameters to the specific values
given in (\ref{E8setup}).
However,
we prolong fixing these parameters until the very end,
anticipating the generalization to the cases with nontrivial
Wilson line parameters.

\subsection{Saddle point equation and resolvent}

Following the example of Nekrasov and Okounkov \cite{Nekrasov:2003rj},
we take the thermodynamic limit of the Nekrasov-type formula
presented in the last section. What we need to do is
to consider the thermodynamic limit $\hbar\to 0$
of the sum over partitions (\ref{Zconformal})
and evaluate the prepotential (\ref{F0inZ}).
This problem has already been solved by
Hollowood, Iqbal and Vafa \cite{Hollowood:2003cv}.
However, in addition to their results
we need the precise form of
the antiderivative of the resolvent
and its analytic properties,
which are in fact essential to our proof.
In this subsection we present a self-contained solution
to the problem with emphasis on the new ingredients.

Let us start with introducing a function $\gamma(z;\hbar)$
which satisfies the difference equation
\[\label{differenceeq}
\gamma(z+\hbar;\hbar)+\gamma(z-\hbar;\hbar)-2\gamma(z;\hbar)
=\ln\varth_1\left(\frac{z}{2\pi}\right)
\]
and has the expansion
\[
\gamma(z;\hbar)=\sum_{g=0}^\infty\hbar^{2g-2}\gamma_g(z).
\]
The explicit form of $\gamma(z;\hbar)$ is not important here.
In the following we merely use the fact that
\[
\gamma_0''(z)=\ln\varth_1\left(\frac{z}{2\pi}\right),
\]
which can be derived immediately
by expanding the above difference equation in $\hbar$.

The summand of the main formula (\ref{Zconformal}) is
expressed as a finite product over boxes in Young diagrams.
It is well known that a product of this kind
can be rewritten as a formally infinite product.
In the present case, (\ref{Zconformal}) is rewritten as
\eqb
\label{Zbis}
Z\Eqn{=}\sum_{\bfR} e^{2\pi i\tvarphi|\bfR|} Z_{\bfR},\nn\\
Z_{\bfR}\Eqn{=}
\prod_{k,l=1}^N
\prod_{\shortstack{\scriptsize$i,j=1$\\
                   \scriptsize$(k,i)\ne(l,j)$}}^\infty
\frac
{\varth_1\left(\tfrac{1}{2\pi}
  (a_k-a_l+(\mu_{k,i}-\mu_{l,j}+j-i)\hbar)\right)}
{\varth_1\left(\tfrac{1}{2\pi}(a_k-a_l+(j-i)\hbar)\right)}\nn\\
&&\times\prod_{k=1}^N \prod_{n=1}^{2N}
\prod_{(i,j)\in R_k}
 \varth_1\left(\tfrac{1}{2\pi}(a_k-m_n+(j-i)\hbar)\right),
\eqe
where
\[\label{tvarphidef}
\tvarphi := \left\{
\begin{array}{ll}
\varphi             &\ \mbox{if $N$ is odd,} \\[1ex]
\varphi+\tfrac{1}{2}&\ \mbox{if $N$ is even.}
\end{array}
\right.
\]
This form is more convenient for our present purposes.
In the thermodynamic limit
the typical size of the partition $\bfR$
contributing to the sum is very large
and $Z$ may be expressed in terms of
continuous profiles of partitions.
Indeed,
using the difference equation (\ref{differenceeq}),
one can verify that $Z_{\bfR}$
is expressed as
\eqb
Z_{\bfR}
\Eqn{=}\exp\left[
-\frac{1}{4}\pint dzdw f''(z)f''(w)\gamma(z-w;\hbar)
 +\frac{1}{2}\sum_{n=1}^{2N}
  \pint dz f''(z)\gamma(z-m_n;\hbar)\right.\nn\\
&&\left.\phantom{\exp\Biggl[}
  +\sum_{k,l=1}^N\gamma(a_k-a_l;\hbar)
  -\sum_{k=1}^N\sum_{n=1}^{2N}\gamma(a_k-m_n;\hbar)\right].
\eqe
Here $f(z)$ is the profile of the partition $\bfR$
\eqb
\label{fdef}
f(z)
\Eqn{=}
\sum_{k=1}^N\Biggl[\sum_{i=1}^{\ell(R_k)}
 \Bigl(|z-a_k-\hbar(\mu_{k,i}-i+1)|
      -|z-a_k-\hbar(\mu_{k,i}-i)|\Bigr)\nn\\
&&\hspace{3em}
      {}+|z-a_k+\hbar\ell(R_k)|\Biggr]
\eqe
and its second derivative is given by
\eqb
\label{ddfdef}
f''(z)\Eqn{=}
2\sum_{k=1}^N\Biggl[\sum_{i=1}^{\ell(R_k)}
 \Bigl(\delta(z-a_k-\hbar(\mu_{k,i}-i+1))
      -\delta(z-a_k-\hbar(\mu_{k,i}-i))\Bigr)\nn\\
&&\hspace{3em}
      {}+\delta(z-a_k+\hbar\ell(R_k))\Biggr]\\
\Eqn{=}
2\sum_{k=1}^N
\Biggl[
\sum_{i=1}^{\infty}
 \Bigl(\delta(z-a_k-\hbar(\mu_{k,i}-i+1))
      -\delta(z-a_k-\hbar(\mu_{k,i}-i))\nn\\
&&\hspace{4.5em}
      {}-\delta(z-a_k+\hbar(i-1))+\delta(z-a_k+\hbar i)\Bigr)
  +\delta(z-a_k)\Biggl].\quad
\eqe
For a partition of large size
$f''(z)$ can be viewed as a density function.
We consider the case where $f''(z)$ has $N$ local supports
respectively around $z=a_k\ (k=1,\ldots,N)$ and
all of them are entirely separated from each other.
We let ${\cal C}_k$ denote the local support
around $z=a_k$ and ${\cal C}$ denote their union,
i.e.~${\cal C}=\cup_{k=1}^N {\cal C}_k$\,.
It follows from the above expression that
\eqb
\label{a_constraints0}
a_k \Eqn{=} \frac{1}{2}\int_{{\cal C}_k}z f''(z) dz,\\
|\bfR|
 \Eqn{=}\frac{1}{4}\int_{\cal C} dz z^2 f''(z)
        -\sum_{k=1}^N\frac{a_k^2}{2}.
\eqe

In the thermodynamic limit,
the sum over partition $Z$ can be approximated by
an integral over the space of continuous functions $f''$
\[\label{Zintegral}
Z\simeq\int {\cal D}f''d^N\lambda \exp
  \left[\frac{1}{2\hbar^2}{\cal F}_0 + {\cal O}(\hbar^0)\right],
\]
where ${\cal F}_0$ is a functional of the following form
\eqb
\label{F0functional}
{\cal F}_0[f'',\lambda_k]
\Eqn{=}-\frac{1}{2}\pint_{\cal C}dzdw f''(z)f''(w)\gamma_0(z-w)
 +\sum_{n=1}^{2N}
  \pint_{\cal C}dz f''(z)\gamma_0(z-m_n)\nn\\
&&{}+4\pi i\tvarphi
  \left(
   \frac{1}{4}\int_{\cal C} dz z^2 f''(z)
  -\sum_{k=1}^N\frac{a_k^2}{2}\right)\nn\\
&&
{}+2\sum_{k=1}^{N}\lambda_k\left(
   \frac{1}{2}\int_{{\cal C}_k}dz z f''(z)-a_k\right).
\eqe
We have introduced Lagrange multipliers $\lambda_k$
taking account of the constraints (\ref{a_constraints0}).
The integral (\ref{Zintegral}) can be evaluated by
the saddle point approximation.
The prepotential (\ref{F0inZ}) is then given,
up to a constant,
by the extremum of the functional ${\cal F}_0$.
Taking the variation of ${\cal F}_0$, one obtains
the saddle point equation
\[\label{saddlepteq}
\pint_{\cal C} dw f''(w)\gamma_0(z-w)
  -\sum_{n=1}^{2N}\gamma_0(z-m_n)-\pi i\tvarphi z^2
  - \lambda_k z 
= 0,
\qquad z\in {\cal C}_k.
\]
Solving this equation with the constraints (\ref{a_constraints0})
and plugging the solution
back into (\ref{F0functional}), one obtains the prepotential $F_0$.

To solve this equation,
it is convenient to consider the following analytic function
\begin{align}
\label{Pidef}
\Omega(z):=
   &\int_{\cal C} f''(w)\gamma_0''(z-w)dw
   -\sum_{n=1}^{2N}\gamma_0''(z-m_n)\nn\\
  =&\int_{\cal C} f''(w)\ln\varth_1\left(\frac{z-w}{2\pi}\right)dw
   - \sum_{n=1}^{2N}\ln\varth_1\left(\frac{z-m_n}{2\pi}\right)
\end{align}
instead of $f''(z)$ itself.
We call it the antiderivative of the resolvent,
as its derivative
\[
\omega(z):=\Omega'(z)
\]
plays the role of the resolvent.
Indeed, the density function $f''$ is recovered as
\[\label{ddfinomega}
2\pi i f''(z)
=\omega(z-i\epsilon)-\omega(z+i\epsilon)
\qquad z\in{\cal C}.
\]
Here $\epsilon=\delta z$ is
an infinitesimal deformation along the cuts,
so that $\pm i\epsilon$ represent infinitesimal deviations
transverse to the cuts.
By definition the Riemann surface of $\Omega(z)$
has logarithmic branches.
It follows that
\[\label{logbranches}
\oint_{\gamma_k}\omega(z)dz=4\pi i,\qquad
\oint_{\gamma^{(n)}}\omega(z)dz=-2\pi i,
\]
where
$\gamma_k\ (k=1,\ldots,N)$ and $\gamma^{(n)}\ (n=1,\ldots,2N)$
denote cycles encircling counterclockwise
the cut ${\cal C}_k$ and the pole at $z=m_n$ respectively.

In terms of $\Omega(z)$, 
the second derivative of the saddle point equation
(\ref{saddlepteq}) is written as
\[
\frac{1}{2}\left(\Omega(z-i\epsilon)+\Omega(z+i\epsilon)\right)
-2\pi i\tvarphi=0\qquad z\in{\cal C}.
\]
Let us solve this equation.
While $\Omega(z)$ has logarithmic branch points
as well as square root branch points,
the following function
\[
G(z):=
  e^{\Omega(z)-2\pi i\tvarphi}
 +e^{-\Omega(z)+2\pi i\tvarphi}
\]
is a meromorphic function on $E$.
Poles at $z=m_n\ (n=1,\ldots,2N)$
are the whole singularities of $G(z)$.
Since (\ref{amrel}) is imposed, $G(z)$ is
strictly doubly periodic.
In other words,
$G(z)$ is an elliptic function of order $2N$.
In terms of $G(z)$, 
$\omega(z)$ is expressed as
\[
\omega(z) = \frac{G'(z)}{\sqrt{(G(z)+2)(G(z)-2)}}.
\]
Since $G(z)\pm 2$ are elliptic functions of order $2N$
and have $2N$ zeros,
the above expression implies that $\omega(z)$ would generically have
$4N$ branch points. On the other hand,
in our setup $\omega(z)$ actually has just
$2N$ branch points. The mismatch is resolved
if the function
\[\label{Hdef}
H(z):=\frac{G(z)+2}{4}=\cosh^2\left(
  \frac{1}{2}\left(\Omega(z)-2\pi i\tvarphi\right)\right)
\]
has $N$ zeros of multiplicity two
instead of $2N$ simple zeros.\footnote{
In general, there are possibilities that $G+2$ and $G-2$
have respectively $N-l$ and $l$ zeros of multiplicity two
($0\le l\le N$).
If parameters are chosen as
$a_k,\hbar\in\bbR$, $\tvarphi,\tau\in i\bbR$ and $m_n=0$,
all ${\cal C}_k$ have to lie on the real axis
and $\exp\Omega(z)>0$ for any $z\in\bbR$
with $z\notin{\cal C}$.
This means that
only the solution with $l=0$ is allowed in this case.
The parameter settings for the $E_r\oplus A_{8-r}\ (r=8,7,6)$
theories are connected with the above setup
by a continuous deformation preserving the topology of the branch
cut configuration of $\Omega(z)$. Thus, the solution with $l=0$
is singled out for these theories.
(The solution with $l=N$ is also allowed, but this is essentially
the same as the solution with $l=0$.)}
The singularities of $H(z)$ are
the single poles at $z=m_n\ (n=1,\ldots,2N)$.
Elliptic functions satisfying these properties are determined as
\[\label{Hsol}
H(z)=\kappa\frac{P(z)^2}{Q(z)}
\]
with
\[\label{generalPQ}
P(z)=\prod_{k=1}^N\varth_1\left(\frac{z-\zeta_k}{2\pi}\right),\qquad
Q(z)=\prod_{n=1}^{2N}\varth_1\left(\frac{z-m_n}{2\pi}\right),
\]
where $\kappa$ and $\zeta_k\ (k=1,\ldots,N)$ are some constants.
The locations of zeros and poles have to satisfy
\[
2\sum_{k=1}^N\zeta_k-\sum_{n=1}^{2N}m_n=0.
\]
Here the equality should be understood modulo periods of the torus $E$.
From (\ref{Hdef})
$\Omega(z)$ is obtained as
\[\label{OmegainH}
\Omega(z)=2\ln\left(\sqrt{H(z)}+\sqrt{H(z)-1}\right)+2\pi i\tvarphi.
\]
By taking the derivative, the resolvent is obtained as
\[
\omega(z)
  =\frac{2\partial_z\sqrt{H(z)}}{\sqrt{H(z)-1}}.
\]
Substituting (\ref{Hsol}) one can verify that this is essentially
equivalent to the resolvent given in \cite{Hollowood:2003cv}.

Recall that $f''(z)$ has to satisfy
the constraints (\ref{a_constraints0}).
In terms of the resolvent, they are expressed as
\[
\label{a_constraints}
a_k=\frac{1}{4\pi i}\oint_{\gamma_k}z\omega(z)dz.
\]
These equations hold if $\omega(z)$ satisfies
\[
\omega(a_k-z\pm i\epsilon)=\omega(a_k+z\pm i\epsilon)
\qquad\mbox{for}\quad a_k+z\in{\cal C}_k.
\]
This holds if the function $H^{1/2}(z):=\sqrt{H(z)}$ satisfies
\[\label{sqrtHoddness}
H^{1/2}(a_k-z)=-H^{1/2}(a_k+z)
\qquad\mbox{for}\quad a_k+z\in{\cal C}_k.
\]
By requiring this property,
the values of $\zeta_k$ are fixed.

Let us now restrict ourselves to the $E_8$ theory
by setting parameters as in (\ref{E8setup}).
In this case, as we will see immediately,
the condition (\ref{sqrtHoddness}) is satisfied
with
\[\label{zetaforE8case}
\zeta_k=\omega_k\qquad (k=1,2,3).
\]
By substituting these data,
the functions $P(z), Q(z)$ are expressed as
\[
P(z)=-iq^{-1/4}
\prod_{k=1}^3\varth_{k+1}\left(\frac{z}{2\pi}\right),\qquad
Q(z)=\varth_1\left(\frac{z}{2\pi}\right)^6.
\]
The function $H=\kappa P^2/Q$ is then obtained as
\[
H(z)
=-\frac{1}{4}u\wp'(z)^2,
\]
where we have used the identity (\ref{wpId3}) and introduced
\[\label{udef}
u:=\frac{4\kappa}{q^{1/2}\eta^{12}}.
\]
Using the property $\wp'(-z)=-\wp'(z)$ and
the periodicity of $\wp'(z)$
one can verify that the above $H(z)$
indeed possesses the property (\ref{sqrtHoddness}).
The resolvent for the $E_8$ theory
is then explicitly expressed as
\[
\label{E8omega}
\omega(z)=
  \frac{2\wp''(z)}{\sqrt{\wp'(z)^2+4u^{-1}}}.
\]
Using (\ref{ddfinomega}) and plugging the above solution back into
(\ref{F0functional}), one obtains the integral expression for
the prepotential.
The Riemann surface of the above resolvent $\omega(z)$ has three cuts
near $z=\omega_k\ (k=1,2,3)$.
The three cuts shrink as $|u|$ increases.
In particular, when $u$ is sent to infinity
all cuts disappear and the Riemann surface of $\omega(z)$ becomes
the torus $E$ with complex modulus $\tau$.
This is reminiscent of the classical limit of
the Seiberg--Witten curve (\ref{E8curve}).
Indeed, the above $u$ is going to
be identified with the coordinate of the Coulomb branch moduli space
in the Seiberg--Witten description.

\subsection{Higgs expectation value and Seiberg--Witten curve}

In this subsection we express $\varphi$
in terms of the function $H(z)$ and reproduce the Seiberg--Witten
description.
To do this, we make use of the following fact
\[\label{intlogtheta}
\frac{1}{2\pi^2 i}
\oint_\alpha\ln\varth_1\left(\frac{z-w}{2\pi}\right)\,dz
= C_1(\tau)\mod\bbZ,
\]
where $C_1(\tau)$ is some function in $\tau$.
The explicit form of $C_1(\tau)$ is not important.
What is crucial here is that $C_1(\tau)$
is independent of $w$ and also 
invariant under continuous deformation
of the integration contour.
This fact can be shown as follows:
Since the theta function is
quasi-periodic $\varth_1(z+1)=-\varth_1(z)$,
function $\frac{1}{2\pi i}\ln\varth_1(\frac{z-w}{2\pi})^2$ is
single-valued modulo $\bbZ$ along a loop belonging
to the cycle $\alpha$. Recall also that the theta function
is regular for $|z|<\infty$,
so that the integral is invariant under the continuous deformation
of the loop.

Substituting (\ref{ddfdef}) into (\ref{Pidef})
and using the above fact one sees that
\[
\frac{1}{4\pi^2 i}\oint_\alpha\Omega(z)dz=0\mod\bbZ,
\]
where $C_1$'s cancel with each other.
Substituting (\ref{OmegainH}), one obtains
\[\label{phiintegralform}
\tvarphi = \frac{i}{2\pi^2}\oint_\alpha
  \ln\left(\sqrt{H(z)}+\sqrt{H(z)-1}\right)dz\mod\bbZ.
\]
This gives an explicit expression of $\varphi$
which is related to $\tvarphi$ by (\ref{tvarphidef}).

We now show that
the above expression is equivalent to the known Seiberg--Witten
description of $\varphi$.
Differentiating the above expression in $u$ one obtains
\[
\frac{\partial\varphi}{\partial u}
  =\frac{i}{4\pi^2 u}\oint_\alpha\frac{dz}{\sqrt{1-H(z)^{-1}}}.
\]
In the case of the $E_8$ theory, it can be written as
\[
\frac{\partial\varphi}{\partial u}
  =\frac{i}{4\pi^2 u}\oint_\alpha
  \frac{\wp'(z)dz}{\sqrt{\wp'(z)^2+4u^{-1}}}.
\]
The Seiberg--Witten curve should be given as
the Riemann surface of the integrand.
It is made of two copies of the torus $E$ connected with
each other by three cuts near $z=\omega_k\ (k=1,2,3)$.
Thus, the Riemann surface is of genus four.
However, by using the identity (\ref{wpId1})
and changing the variables as
\[\label{fromztox}
\wp(z)=u^{-2}x,
\]
one obtains
\[\label{dphidu}
\frac{\partial\varphi}{\partial u}
=\frac{i}{4\pi^2}\oint_\talpha \frac{dx}{y},
\]
where $y$ is given by
\[\label{E8curvebis}
y^2=4x^3-\frac{1}{12}E_4 u^4x-\frac{1}{216}E_6 u^6+4u^5.
\]
This is exactly the
Seiberg--Witten curve (\ref{E8curve}) for the E-string theory!
It is clear from the definitions
(\ref{talphatbeta}), (\ref{alphabeta}) that
$\talpha$ is the image of $\alpha$ by the map (\ref{fromztox}). 
Thus, the above expression for $\varphi$ is in perfect agreement
with the Seiberg--Witten description of the Higgs expectation value
(\ref{vevs}).

\subsection{Dual Higgs expectation value and prepotential}

To complete our proof, we need to show that the prepotential obtained
from the Nekrasov-type formula is also expressed in terms of
period integrals as in (\ref{dF0inphi}) with (\ref{vevs}).
For this purpose
we consider the contour integral of $\Omega(z)$ around
the cycle $\beta$. To do this, we make use of
the modular transformation law of the theta function
\[
\varth_1\left(\frac{z}{2\pi},\tau\right)
=e^{3\pi i/4}
\tau^{-1/2}\exp\left(-\frac{i z^2}{4\pi\tau}\right)
  \varth_1\left(\frac{z}{2\pi\tau},-\frac{1}{\tau}\right).
\]
Using this and
applying (\ref{intlogtheta}) with modulus $-1/\tau$,
one can show that
\eqb\label{intlogthetabeta}
\lefteqn{
\frac{1}{2\pi^2 i\tau}
\int_{z_0}^{z_0+2\pi\tau}
  \ln\varth_1\left(\frac{z-w}{2\pi},\tau\right)\,dz}\nn\\
\Eqn{=}-\frac{1}{8\pi^3\tau^2}\int_{z_0}^{z_0+2\pi\tau} (z-w)^2 dz\nn\\
&&
+\frac{1}{2\pi^2 i\tau}\int_{z_0}^{z_0+2\pi\tau}
  \ln\varth_1\left(\frac{z-w}{2\pi\tau},-\frac{1}{\tau}\right)\,dz
+\frac{3}{4}-\frac{1}{2\pi i}\ln\tau
\nn\\
\Eqn{=}-\frac{1}{8\pi^3\tau^2}\int_{z_0}^{z_0+2\pi\tau} (z-w)^2 dz
+ C_1\left(-\frac{1}{\tau}\right)
+\frac{3}{4}-\frac{1}{2\pi i}\ln\tau
\mod\bbZ\nn\\
\Eqn{=}-\frac{1}{4\pi^2\tau}w^2
  +\left(\frac{1}{2\pi}+\frac{z_0}{2\pi^2\tau}\right) w
+ C_2(z_0,\tau) \mod\bbZ,\quad
\eqe
where $C_2(z_0,\tau)$ is some function in $z_0$ and $\tau$.
By using this together with (\ref{ddfdef}) and (\ref{Pidef}),
one obtains
\eqb
\lefteqn{
\frac{1}{4\pi^2 i\tau}\int_{z_0}^{z_0+2\pi\tau}\Omega(z)dz}\nn\\
\Eqn{=}
-\frac{1}{8\pi^2\tau}\pint_{\cal C} w^2 f''(w) dw
+\left(\frac{1}{4\pi}+\frac{z_0}{4\pi^2\tau}\right)
\pint_{\cal C} w f''(w) dw \mod\bbZ\nn\\
\Eqn{=}\frac{i}{8\pi^3\tau}\left(\frac{\partial F_0}{\partial \varphi}
  +2\pi i\sum_{k=1}^N a_k^2\right)
  +\left(\frac{1}{2\pi}+\frac{z_0}{2\pi^2\tau}\right)\sum_{k=1}^N a_k
  \mod\bbZ.
\label{openbetaint}
\eqe
$C_2$'s cancel with each other in the first equality.
To show the second equality
we use (\ref{a_constraints0})
and (\ref{F0functional})
with
\[
\frac{\partial F_0}{\partial \varphi}
=\frac{\partial {\cal F}_0}{\partial \varphi}\bigg|_{\rm extremum}
=\left[\left(
\frac{\partial {\cal F}_0}{\partial \varphi}\right)_{f''}
+\left(\frac{\delta {\cal F}_0}{\delta f''}\right)_\varphi
 \frac{\partial f''}{\partial \varphi}\right]_{\rm extremum}
=\left[\left(
\frac{\partial {\cal F}_0}{\partial \varphi}
\right)_{f''}\right]_{\rm extremum}.
\]
Here $(\partial {\cal F}_0/\partial \varphi)_{f''}$
denotes the partial derivative of ${\cal F}_0$
with respect to $\varphi$, holding $f''$ constant.

Let us now restrict ourselves to the $E_8$ theory.
In this case the second term of the last line in (\ref{openbetaint})
vanishes as we set $a_k=\omega_k\ (k=1,2,3)$.\footnote{
The second term in (\ref{openbetaint}) actually vanishes
not only in the $E_8$ case but in most of
the cases with generic four Wilson line parameters
(\ref{fourWsetup}). On the other hand, it does not vanish
in some special cases, such as the case of
$E_7\oplus A_1$ symmetry, which we will study later. In these cases,
the contour integral of $\Omega(z)$ along the cycle $\beta$
makes sense only up to a constant.
Anyway, we will eventually show a relation
up to a constant and thus such a constant ambiguity is irrelevant.}
Thus, the integral
is actually independent of $z_0$ and is regarded as
the period integral over the cycle $\beta$.
To sum up, one obtains
\[
\frac{1}{4\pi^2 i\tau}\oint_\beta\Omega(z)dz
=\frac{i}{8\pi^3\tau}\left(\frac{\partial F_0}{\partial \varphi}
  +2\pi i\sum_{k=1}^3 \omega_k^2\right)\mod\bbZ.
\]
On the other hand, by using (\ref{OmegainH}) the same period integral
is expressed as
\eqb
\frac{1}{4\pi^2 i\tau}\oint_\beta\Omega(z)dz
\Eqn{=}\frac{1}{2\pi^2 i\tau}\oint_\beta
  \ln\left(\sqrt{H(z)}+\sqrt{H(z)-1}\right)dz
+\tvarphi\nn\\
\Eqn{=}-\frac{1}{\tau}\varphi_{\rm D}+\varphi+\mbox{const.},
\eqe
where we have identified
the dual Higgs expectation value $\varphi_{\rm D}$ as
\[
\varphi_{\rm D}= \frac{i}{2\pi^2}\oint_\beta
  \ln\left(\sqrt{H(z)}+\sqrt{H(z)-1}\right)dz+\mbox{const.}
\]
regarding the expression (\ref{phiintegralform}) for $\varphi$.
Here const.'s are some functions in $\tau$ but are
independent of $\varphi$.
By comparing these two expressions, one obtains
\[
\frac{\partial F_0}{\partial \varphi}
 =8\pi^3i\left(\varphi_{\rm D}-\tau\varphi\right)+\mbox{const.}
\]
This is in perfect agreement with the Seiberg--Witten description 
(\ref{dF0inphi}).

\section{Cases with other global symmetries}

The proof presented in the last section can be generalized to the cases
with nontrivial Wilson line parameters. As an illustration, we briefly
discuss two examples which have
$E_7\oplus A_1$ or $E_6\oplus A_2$ as a global symmetry.

\subsection{$E_7\oplus A_1$ theory}

The case of $E_7\oplus A_1$ global symmetry
is realized by setting the parameters as
\[
N=2,\qquad
a_k=\omega_{k+1}\quad (k=1,2),\qquad
 m_n=0\quad (n=1,2,3,4).
\]
In this case, the condition (\ref{sqrtHoddness}) is satisfied
if we choose $\zeta_k$ as
\[
\zeta_k=\omega_{k+1}\quad (k=1,2).
\]
The function $H(z)$ is obtained as
\[
H(z)
=\frac{u\varth_3^2\varth_4^2}{16}
\frac{\wp'(z)^2}{\wp(z)-e_1},
\]
where $u$ is defined as in (\ref{udef}) but with an opposite sign
and
\[
e_1=\frac{\varth_3^4+\varth_4^4}{12}.
\]
We have abbreviated $\varth_k(0,\tau)$ as $\varth_k$.
Let us introduce a new variable $x$ by
\[
\wp(z)-e_1=u^{-2}x.
\]
One then obtains
\[
\frac{\partial\varphi}{\partial u}
=\frac{i}{4\pi^2}\oint_\talpha \frac{dx}{y},
\]
where $y$ is given by
\[
y^2=4x^3+\left(\varth_3^4+\varth_4^4\right)u^2x^2
  +\left(\frac{\varth_3^4\varth_4^4}{4}u
    -\frac{16}{\varth_3^2\varth_4^2}\right)u^3x.
\]
This is exactly the Seiberg--Witten curve
for the $E_7\oplus A_1$ case \cite{Sakai:2012ik}.

\subsection{$E_6\oplus A_2$ theory}

The case of $E_6\oplus A_2$ global symmetry
is realized by setting the parameters as
\[
N=3,\qquad
a_k=\omega_k\quad (k=1,2,3),\qquad
m_n=-m_{n+3}=\frac{2\pi}{3}\quad (n=1,2,3).
\]
In this case, the condition (\ref{sqrtHoddness}) is satisfied
if we choose $\zeta_k$ as
\[
\zeta_k=\omega_k\quad (k=1,2,3).
\]
Let us introduce the notation
\[
\alpha_3:=
  \varth_3(0,2\tau)\varth_3(0,6\tau)
 +\varth_2(0,2\tau)\varth_2(0,6\tau),\qquad
\beta_3:=\frac{\eta(\tau)^9}{\eta(3\tau)^3}.
\]
By using the following identity
\[
\varth_1\left(\frac{z}{2\pi}-\frac{1}{3}\right)
\varth_1\left(\frac{z}{2\pi}+\frac{1}{3}\right)
=-3\frac{\eta(3\tau)^2}{\eta(\tau)^6}
  \left(\wp(z)-\frac{1}{4}\alpha_3^2\right)
  \varth_1\left(\frac{z}{2\pi}\right)^2
\]
and (\ref{wpId3}), the function $H(z)$ is obtained as
\[
H(z)
 =\frac{u\beta_3^2}{108}
  \frac{\wp'(z)^2}{\left(\wp(z)-\frac{1}{4}\alpha_3^2\right)^3}.
\]
Here $u$ is given by (\ref{udef}).
Let us introduce a new variable $x$ by
\[
\wp(z)-\frac{1}{4}\alpha_3^2=\frac{x}{u(u-27\beta_3^{-2})}.
\]
One then obtains
\[
\frac{\partial\varphi}{\partial u}
=\frac{i}{4\pi^2}\oint_\talpha \frac{dx}{y},
\]
where $y$ is given by
\[
y^2=4x^3+3\alpha_3^2u^2x^2
  +\frac{2}{3}\alpha_3\left(\beta_3u-\frac{27}{\beta_3}\right)u^3x
  +\frac{1}{27}\left(\beta_3u-\frac{27}{\beta_3}\right)^2u^4.
\]
This is exactly the Seiberg--Witten curve
for the $E_6\oplus A_2$ case \cite{Sakai:2012ik}.

\section{Discussion}

We have proved the Nekrasov-type formula for the Seiberg--Witten
prepotential for the E-string theory on $\bbR^4\times T^2$.
Following the example of Nekrasov--Okounkov, we have taken
the thermodynamic limit of the Nekrasov-type formula and
have determined the profile which dominates the saddle point
approximation of the sum over partitions.
Due to the difference in identification of parameters between the
E-string theory and ordinary gauge theories, the proof
by Nekrasov--Okounkov cannot be straightforwardly generalized. We have
resolved this problem by considering the antiderivative of the
resolvent rather than the resolvent itself in the thermodynamic limit.

The Seiberg--Witten curve obtained directly from the Nekrasov-type
formula is of genus greater than one and is not an elliptic curve.
We have found a simple transformation of variables
by means of the Weierstrass $\wp$-function
which maps the higher genus curve to the known elliptic
Seiberg--Witten curve for the E-string theory.
Such a simplification is possible
because the parameters
in the Nekrasov-type formula have been chosen specifically
for the setup of the E-string theory.

As the E-string theory is one of the simplest
non-Lagrangian field theories, the theory is
ubiquitous in the study
of such theories in higher dimensions. For instance,
the five-dimensional limit of the E-string theory with $E_6$
global symmetry is identical to the $T_3$ theory in five dimensions 
\cite{Benini:2009gi}, for which Nekrasov-type partition functions
have been studied recently \cite{Bao:2013pwa,Hayashi:2013qwa}.
We hope that investigations into Nekrasov-type formulas
for the E-string theory will shed light on
the mysterious nature of non-Lagrangian field theories.

\vspace{3ex}

\begin{center}
  {\bf Acknowledgments}
\end{center}

The authors would like to thank T.~Eguchi,
S.~Moriyama, K.~Ohta and especially Y.~Sugawara
for helpful comments and discussions.
The work of K.S.~is supported in part by Grant-in-Aid
for Scientific Research from the Ministry of Education, Culture, 
Sports, Science and Technology of Japan (MEXT).

\vspace{3ex}


\appendix

\section{Conventions of special functions}

The Jacobi theta functions are defined as
\eqb
\varth_1(z,\tau)\Eqn{:=}
   i\sum_{n\in \bbZ} (-1)^n y^{n-1/2}q^{(n-1/2)^2/2},\\
\varth_2(z,\tau)\Eqn{:=}
   \sum_{n\in \bbZ} y^{n-1/2}q^{(n-1/2)^2/2},\\
\varth_3(z,\tau)\Eqn{:=}
   \sum_{n\in \bbZ} y^n q^{n^2/2},\\
\varth_4(z,\tau)\Eqn{:=}
   \sum_{n\in \bbZ} (-1)^n y^n q^{n^2/2},
\eqe
where $y=e^{2\pi i z},\ q=e^{2\pi i \tau}$. We often use the following
abbreviated notation
\[
\varth_k(z) := \varth_k(z,\tau),\qquad
\varth_k := \varth_k(0,\tau).
\]
The Dedekind eta function is defined as
\[
\eta(\tau) := q^{1/24}\prod_{n=1}^\infty (1-q^n).
\]
The Eisenstein series are given by
\[
E_{2n}(\tau)
   =1+\frac{2}{\zeta(1-2n)}
   \sum_{k=1}^{\infty}\frac{k^{2n-1}q^k}{1-q^k}.
\]
We often abbreviate $\eta(\tau),\,E_{2n}(\tau)$ as $\eta,\,E_{2n}$
respectively.

The Weierstrass $\wp$-function is defined as
\[
\wp(z)=\wp(z;2\omega_1,2\omega_3)
  :=\frac{1}{z^2}
  +\sum_{(m,n)\in\bbZ^2_{\ne (0,0)}}
  \left[\frac{1}{(z-\Omega_{m,n})^2}
    -\frac{1}{{\Omega_{m,n}}^2}\right],
\]
where $\Omega_{m,n}=2m\omega_1 + 2n\omega_3$. We also introduce
the following notation
\[
e_k := \wp(\omega_k)\qquad (k=1,2,3),
\]
with
\[
\omega_1+\omega_2+\omega_3 = 0,\qquad
\frac{\omega_3}{\omega_1} = \tau.
\]
In the main text we use the following identities
\eqb
\wp'(z)^2
\label{wpId1}
\Eqn{=}4\wp(z)^3
  -\frac{\pi^4}{12\omega_1^4}E_4\wp(z)
  -\frac{\pi^6}{216\omega_1^6}E_6\\
\label{wpId2}
\Eqn{=}4(\wp(z)-e_1)(\wp(z)-e_2)(\wp(z)-e_3)\\
\Eqn{=}
\label{wpId3}
\frac{\pi^6}{\omega_1^6}\eta^{12}
  \prod_{k=1}^3
  \frac{\varth_{k+1}(\frac{z}{2\pi})^2}{\varth_1(\frac{z}{2\pi})^2}.
\eqe
%


\renewcommand{\section}{\subsection}
\renewcommand{\refname}

\end{document}